\documentclass[amsmath, amsfonts, twocolumn, superscriptaddress,prx]{revtex4-1}
\usepackage{graphicx}
\usepackage{epsfig}
\usepackage{bm}
\usepackage{dcolumn}
\usepackage{amsmath}
\usepackage{amssymb}
\usepackage{color}

\newcommand{\dv}{\mathop{\rm div}\nolimits}
\newcommand{\Li}{\mathop{\rm Li}\nolimits}

\begin{document}

\title{Magnetodrag in the hydrodynamic regime: \\ Effects of magnetoplasmon resonance and Hall viscosity}
 
\author{S. S. Apostolov} 
\affiliation{A. Ya. Usikov Institute for Radiophysics and Electronics NASU, 61085 Kharkov, Ukraine}

\author{D. A. Pesin}
\affiliation{Department of Physics, University of Virginia, Charlottesville, VA 22904, USA}

\author{A. Levchenko}
\affiliation{Department of Physics, University of Wisconsin--Madison, Madison, Wisconsin 53706, USA}

\begin{abstract}
In this work we study magnetotransport properties in electronic double layers of strongly correlated electron liquids. For sufficiently clean high-mobility samples, the high-temperature regime of transport in these systems can be described in pure hydrodynamic terms. We concentrate on the magnetic field dependence of longitudinal drag effect mediated by the interlayer Coulomb scattering and identify several mechanisms of transresistance which is caused by viscous flows, magnetoplasmon resonance, and dissipative thermal fluxes. In particular, we elucidate how Hall viscosity enters magnetodrag and modifies its temperature dependence in the magnetic field. 
\end{abstract}

\date{July 11, 2019}

\maketitle

\section{Introduction and motivation} 

Electronic transport experiments in heterostructures consisting of two spatially separated conducting layers provide a versatile platform to directly study 
manifestations of electronic correlations promoted by the long-range nature of Coulomb interaction. The canonical example is the Coulomb drag setup 
of weakly coupled two-dimensional layers, wherein interlayer electron scattering gives raise to a nonlocal response, namely a voltage drop $V_{\mathrm{drag}}$ 
that appears in one of the layers due to the electrical current $I_{\mathrm{drive}}$ flowing in the other \cite{Drag-Review}. The resulting drag resistivity, $\rho_D=V_{\mathrm{drag}}/I_{\mathrm{drive}}$ \cite{Note1}, is extremely susceptible to temperature ($T$), magnetic field ($H$), interlayer spacing ($d$), intralayer carrier density ($n$) or density mismatch between the layers ($\delta n$), and intralayer mean free path ($l$), which can be dominated by either impurity scattering, in the disordered case, or by inter-electron collisions in clean systems. Importantly, the transresistivity reflects not only the exact character of interlayer interaction, but also the nature of elementary excitations in each layer and thus gives practical tools to elucidate their fundamental properties. This becomes especially pronounced in the regime of strongly coupled layers at small layer separation where drag reveals superfluidity of indirect excitons \cite{Nandi} and broken symmetries in the double-layered systems of monolayer and bilayer graphene \cite{Geim,Kim,Dean,Hone}. It has been demonstrated recently that the extent of many-body correlations can be further enriched by incorporating a twist angle between the layers \cite{PJH}.  
  
In this work we study magnetodrag between quantum wells of strongly correlated electrons at large interlayer separations. In this case the interlayer coupling is weak and can be treated perturbatively, yet the intralayer interactions should be accounted for exactly. More specifically, we consider a high-mobility and low-density two-dimensional (2D) electron systems in a situation when the intralayer interaction parameter, so-called Wigner-Seitz dimensionless radius $r_s$ that is understood as the ratio of the average potential energy to the average kinetic energy, $r_s\simeq (na^2_B)^{-1/2}\sim e^2/(\epsilon v_F)$, can be large, $1< r_s< r^c_s$. Here $a_B=\epsilon/m e^2$ is the effective Bohr radius in the material with the dielectric constant $\epsilon$ and the effective mass $m$. The critical value of $r^c_s$ marks the onset of Wigner crystallization in 2D and is estimated to be of the order $r^c_s\sim 40$ \cite{Tanatar,Attaccalite}. For sufficiently clean samples at large $r_s$ and at moderately high temperatures, the electron-electron scattering length $l_{\mathrm{ee}}$ can become short compared to other scales at which electrons relax their momentum by electron-impurity or interlayer electron-electron scattering. If, in addition, the temperature is not too high, such that the energy- and momentum-relaxing electron-phonon scattering can be neglected, the electronic liquid exhibits hydrodynamic behavior. In this scenario, the complexities of intralayer interactions can be described by only a few kinetic coefficients, such as viscosity and thermal conductivity, in the framework of Navier-Stokes equation of a charged fluid \cite{Andreev,AL-Viscous-MR}. 
  
There are several key factors stemming from both experimental observations and theoretical proposals that motivate this work. It has been shown early on that in the collisionless limit with respect to intralayer scattering and to the leading second-order in interlayer interaction, longitudinal magnetodrag is absent (at least to $H^2$ order), as all field dependent terms cancel out upon inverting the drag conductivity matrix, and Hall drag vanishes as well \cite{Kamenev}. These results hold unless some special additional assumptions are made concerning the energy dependence of the quasiparticle scattering time that may lift cancellations and lead to weak Hall drag signal \cite{Hu}. On the other hand, experiments show strong magnetodrag resistance with pronounced temperature dependence $\rho_D\propto T H^2$ \cite{Hill-1,Hill-2}. Furthermore, the magnitude of drag effect increases sharply with large $r_s$ \cite{Pillarisetty-1,Pillarisetty-2,Spivak}. This seems to suggest that intralayer interaction processes are of principal importance and can not be disregarded. The situation with the Hall drag is more subtle since there is only a very limited data available in the weak field. Most of the experiments have been conducted in high quantizing fields at low temperatures in Shubnikov-de Haas and integer quantum Hall regimes \cite{Rubel,Gramila,Lilly,Kellogg-1,Kellogg-2,Muraki} where the sign of magnetodrag depends on Landau level filling factor difference between the two layers. This limit is well described by existing theories \cite{Bonsager,Khaetskii,Oppen,Manolescua,Badalyan,Gornyi,Metzner}. 

The hydrodynamic regime of drag effect has recently attracted considerable attention, and have been studied theoretically in the context of strongly correlated electronic double layers \cite{Apostolov,Chen,Patel,Holder-Drag} and graphene layers close to neutrality \cite{Song,Abanin,Titov,Narozhny}. The magnetodrag -- the drag effect in an external magnetic field -- has not been studied in the context of hydrodynamic transport in strongly interacting electronic fluids at large $r_s$. In this paper, we study the Coulomb drag effect in the presence of an out-of-plane magnetic field in the hydrodynamic regime. We show that the magnetodrag is non-zero even in classical magnetic fields (Landau quantization not being important). The main effects of the magnetic field are to modify the spectrum of collective modes (plasmons); change the transport coefficients of the liquid, in particular its viscosity tensor and thermal conductivity. 

One particularly interesting aspect of the electronic magnetohydrodynamics is the appearance of the Hall viscosity --  an antisymmetric component of the viscosity tensor in a magnetic field. The Hall viscosity has been a subject of an intense theoretical scrutiny over the years, as it is related to a topological property of the quantum Hall state \cite{Avron,Haldane,Son,Read}. There have been several  proposals on how to measure its classical counterpart via bulk magnetoresistance in planar and Corbino geometries \cite{Polini,Moore,Gromov,Holder-Corbino,Hankiewicz}. However, measurements of Hall viscosity have been realized only recently in graphene electron liquid as reported by the Manchester group \cite{Berdyugin}. Here, we explore how the Hall viscosity enters the magnetodrag resistance. While our results are limited to a classically strong magnetic field field, qualitatively they should be applicable as long as temperature is high enough that thermal broadening effects smear out Landau quantization. A crossover from semiclassical to Shubnikov--de Haas regime was considered recently in the context of Hall viscosity calculation \cite{Burmistrov}. In principle, a similar approach can be used in application to the Coulomb drag problem.      

The remainder of the paper is organized as follows. First, in Section \ref{sec:qualitative} we present our results at the qualitative level and explain their physical origin since some of the expressions we find can be anticipated based on a very general considerations. The next central Section \ref{sec:technique} of the paper contains technical details of our calculations which are based on the stochastic magnetohydrodynamic theory of the Navier-Stokes equation with Langevin forces. In the last Section \ref{sec:discussion} we discuss emergent transport regimes, provide estimates of the relative importance of various terms in drag magnetoresistance, and place our findings in the context of published experiments. In the Appendices we show that a modified structure of the stress tensor in the field does not translate into the correlation function of Lagnevin fluxes, and calculate asymptotes of certain integrals. In conclusions, we also briefly comment on possible further developments and remaining outstanding questions.   

\section{Qualitative considerations} \label{sec:qualitative}

\subsection{Temperature regimes}

We start with a discussion of temperature regimes that limit the applicability of the expressions obtained in this paper. 

For weak impurity and phonon scattering, the onset of collision-dominated regime in a bilayer is set by the temperature $T_{\mathrm{col}}$ at which the intralayer electron mean free path becomes comparable to the interlayer separation, $l_{\mathrm{ee}}\lesssim d$. A specific estimate for $T_{\mathrm{col}}$ depends on the nature of quasiparticle scattering. Throughout the paper, we assume that the Fermi-liquid (FL) picture applies, in which $l_{\mathrm{ee}}=v_F\tau_{\mathrm{ee}}\propto v_FE_F/T^2$. There is direct experimental evidence that this relationship holds for the strongly interacting regime $r_s\sim 10$ \cite{Eisenstein}. In this regime, we deduce $T_{\mathrm{col}}\sim E_F/\sqrt{k_Fd}$. In our analysis we systematically assume that $k_Fd>1$, which is a valid regime for most of the double layered systems in drag experiments, except for double layers with graphene where this condition can be reversed at low doping. 

In this high-temperature regime, $T>T_{\mathrm{col}}$, one must take into account intralayer electron-electron collisions while considering the interlayer drag effect. A fully hydrodynamic description of the drag effect is possible at yet higher temperatures, the scale for which is set by comparing the frequency of the two-dimensional plasmon mode $\omega_{\rm pl}$ (see Eq.~\ref{md-rho-pl} and the text below it) with the wave vector $q\sim 1/d$, to the inverse electron-electron intralayer scattering time. For $\omega_{\rm pl}\lesssim 1/\tau_{\rm ee}$, transport in an electronic double-layer is dominated by hydrodynamic density fluctuations. Besides the plasmon, the latter include propagating diffusive modes of thermally expanding electron liquid caused by temperature fluctuations. It is easy to show that the fully hydrodynamic regime sets in at temperatures above $T_{\rm hydro}\sim E_F/\sqrt[4]{k_Fd}$. 

The energy scale that sets the onset of strong electron-phonon scattering is determined by the Bloch-Gr\"uneisen temperature, $T_{\mathrm{ph}}=2k_Fu_s$, where $u_s$ is the speed of sound. It is important that for 2D systems it is possible to reach a regime where $T_{\mathrm{ph}}>E_F$. Indeed, Fermi energy scales linearly with electron density, $E_F\propto n$, whereas $T_{\mathrm{ph}}\propto\sqrt{n}$, so that for semiconductor heterostructures at sufficiently low carrier densities $T_{\mathrm{ph}}$ can be above $E_F$. This condition is fulfilled in numerous experiments [see Ref. \cite{Gao} for a review of experimental data and a related discussion]. At temperatures below $T_{\mathrm{ph}}$, the electron-phonon scattering time is long as it scales as a high power of temperature $\tau^{-1}_{\mathrm{e-ph}}\simeq \xi_{\mathrm{ph}} (T/T_{\mathrm{ph}})^5$, where energy scale of $\xi_\mathrm{ph}$ is proportional to the square of the deformation potential \cite{Price}. This behavior guarantees that $\tau_{\mathrm{ee}}\ll\tau_{\mathrm{e-ph}}$. Lastly, in sufficiently clean samples, it is possible to ensure that the electron-impurity scattering rate is small compared to the momentum exchange rate between the layers of the system. Under such conditions, impurity scattering can be neglected while considering the drag effect. 

In conclusion, we assume the following hierarchy of energy scales $T_{\mathrm{col}}<T_{\mathrm{hydro}}<E_F<T_\mathrm{ph}$ and the expressions for magnetodrag resistance obtained in this work apply for $T_{\rm hydro}<T<T_\mathrm{ph}$.

\subsection{Mechanisms of magnetodrag}

To begin the discussion of magnetodrag mechanisms, we recall the scale of the drag resistance in the hydrodynamic regime \cite{Apostolov}:
\begin{align}\label{eq:rho_d}
\rho_D\simeq \frac{\rho_Q}{(k_Fd)^4}\frac{a_B\eta}{nd}\frac{T}{E_F}.
\end{align}
Here $\rho_Q=2\pi/e^2$ is the resistance quantum, and $\eta$ is the shear viscosity of electron fluid. In this formula, and all the forthcoming estimates in this section, we omitted all numerical coefficients and cumbersome logarithmic in temperature renormalization factors, which we derive explicitly later in the text. The expressions for $\rho_D$ in Eq. \eqref{eq:rho_d} can be understood from the following physical  considerations. In the collision-dominated regime, fluctuations of the viscous stresses that excite density modes can be described by stochastic Langevin forces in the Navier-Stokes equation. The latter obey Gaussian distribution and the fluctuation-dissipation theorem dictates that they correlate to corresponding kinetic coefficient, which is the shear viscosity, $\eta$. Since drag is governed by the coupling of density modes we expect that temperature dependence of corresponding resistance is determined by the product $T\eta$. To understand density and spacing dependence of $\rho_D$, we first realize that the typical momentum transfer between the layers is set by the inverse layer separation, $q\sim 1/d$, which is basically a property of screened Coulomb interaction $\propto (e^2/q)\mathrm{e}^{-qd}$. To the linear order in the electron flow velocity $\bm{v}$ of the drive layer the drag force $\bm{F}_D$ is proportional to a product $\bm{q}(\bm{q}\cdot\bm{v})$. As one needs to sum over all transferred momenta, the phase space in 2D brings a factor $q^2$, the correlation function of fluctuating stresses gives another $q^4$ (this is because stress tensor contains gradient of velocity, and another gradient in the Navier-Stokes equation that needs to be squared in the correlation function), and lastly dynamical screening effects associated plasmon modes bring another $1/q^2$. All together this gives $F_D\propto 1/d^5$ while the remaining density dependence can be restored from the dimensional argument recalling that shear viscosity has units of density so that $F_D\propto 1/n^4$ leading to Eq. \eqref{eq:rho_d}.

The first mechanism of magnetodrag comes from the magnetic field dependence of the shear viscosity: in a finite field electrons experience cyclotron motion, which suppresses the viscosity, $\eta\to \eta(H)$, by shortening the mean free path in the direction of flow \cite{Steinberg}:
\begin{equation}\label{eta-xx}
\eta(H)=\eta/[1+(2\omega_c\tau_{\mathrm{ee}})^2].
\end{equation}
Here $\omega_c=eH/mc$ is the cyclotron frequency. For the drag magnetoresistance, $\delta\rho_D(H,T)=\rho_D(H,T)-\rho_D(0,T)$, this gives rise to a negative viscous drag correction $\delta \rho^{\mathrm{visc}}_D$ in the weak field limit, $\omega_c\tau_{\mathrm{ee}}\ll1$:
\begin{equation}\label{md-rho-visc}
\delta\rho^{\mathrm{visc}}_D\simeq -\rho_D (\omega_c\tau_{\mathrm{ee}})^2.
\end{equation}
              
In order to explain the second contribution to drag resistance, we recall that the plasmon dispersion spectrum $\omega_{\pm}(q)$ for a double 2D electron system consists of two branches of excitations. These are the acoustic (optical) plasmon modes, where the charge density oscillations in the two layers occur in antiphase (phase). It should be noted that in the hydrodynamic limit the imaginary part of the plasmon dispersion is determined by electron viscosity, $\Im \omega_{\pm}\propto \eta q^2/(mn)$, so that fluctuations with sufficiently low wave numbers $q$ have large mean free paths, and therefore plasmons are well-defined excitations in this regime. As explained in Ref.  \cite{Hruska} this is also the physical reason for fluctuational mechanism of viscosity to be the basic one at sufficiently small gradients and in samples with no elastically scattering potential. At the finite field the plasmon dispersion is pushed to higher energies, $\omega_\pm(H,q)=\sqrt{\omega^2_\pm(q)+\omega^2_c}$, so that it is harder to excite plasmon resonance and correspondingly drag friction decreases as 
 \begin{equation}\label{md-rho-pl}
 \delta\rho^{\mathrm{pl}}_D\simeq-\rho_D(\omega_c/\omega_{\mathrm{pl}})^2,\quad \omega_{\mathrm{pl}}=\sqrt{\frac{2\pi e^2n}{\epsilon md}}
 \end{equation}
 where $\omega_{\mathrm{pl}}$ is the characteristic plasmon energy computed at the typical wave number of interlayer coupling.
 
Yet another  - third - contribution to the magnetodrag in the hydrodynamic regime comes from the modification of the plasmon dispersion by the Hall viscosity, $\eta_{\mathrm{H}}$. This effect has been recently considered in the context of edge magnetoplasmon excitations \cite{Goldstein}, and in the context of magnetic resonance in a high-frequency flow of a viscous electron fluid \cite{Alekseev}. We consider this effect in a bulk of a bilayer system. Hall viscosity results in the contribution to magnetic transresistance of the form  
 \begin{equation}\label{md-rho-Hall-visc}
\delta\rho^{\mathrm{H}}_D\simeq -\rho_D(\omega_c/\omega_{\mathrm{pl}})^2(l^2_H\eta_{\mathrm{H}}/nd^2).
\end{equation}
We recall that the classical Hall viscosity \cite{Steinberg}
\begin{equation}\label{eta-xy}
\eta_{\mathrm{H}}=\eta (2\omega_c\tau_{\mathrm{ee}})/[1+(2\omega_c\tau_{\mathrm{ee}})^2], 
\end{equation}
is linear in $H$ at weak fields so its product with the square of magnetic length $l_H=\sqrt{c/eH}$ is independent of the field in the main approximation. Furthermore, by using an electron gas-kinetic formula for viscosity, $\eta=mv_Fnl_{\mathrm{ee}}$ it is easy to show that the term in the last brackets of Eq. \eqref{md-rho-Hall-visc} is parametrically equal to the square of the Knudsen number $\mathrm{Kn}=l_{\mathrm{ee}}/d$. 
  
The forth and the final contribution to the magnetodrag that we have identified is an energy driven mechanism related to the propagating neutral modes. Indeed, thermal fluctuations drive diffusive energy fluxes through thermal conduction that locally heat electron fluid. These processes can be accounted by introducing an additional Langevin forces that lead to thermal expansion and thus change of electron density. The corresponding entropy production gives another channel of dissipation and enhances interlayer friction via plasmons  
\begin{equation}\label{md-rho-th}
\delta\rho^{\mathrm{th}}_{D}\simeq (\rho_Q/nd^2)(c_VT/\kappa)(u_T/d\omega_{\mathrm{pl}})^4(\omega_c\tau_{\mathrm{ee}})^2.
\end{equation}
Here $\kappa$ is the electronic thermal conductivity, $c_V=T(\partial s/\partial T)_V$ is specific heat per particle,  and $u_T=\sqrt{T/mc_v}(\partial s/\partial \ln n)_T$ is the thermal mode velocity. Unlike the viscous mechanism, thermal drag is enhanced by magnetic field. 

The fact that $\delta\rho^{\mathrm{th}}_{D}$ is inversely proportional to $\kappa$ is rather subtle and deserves a brief explanation. In analogy with the viscous term, the thermal Langevin forces correlate to thermal conductivity by fluctuation-dissipation relation so that the corresponding strength of density response is proportional to $T\kappa$. However, unlike the density modes of plasmon excitation, neutral thermal mode is purely diffusive with overdamped dispersion relation $\omega=\mathrm{i}\kappa q^2/nc_V$. A slow thermal diffusion effectively anti-screens dynamical Coulomb interaction of long wavelength density fluctuations which scales as $1/(\kappa q^2)^2$. This mechanism is similar to that of drag enhancement in the disordered case as screening is less effective for diffusive modes \cite{Kamenev}. It is also analogous to the previous discussion of drag enhancement by emergent magnetoviscous modes. Lastly, it should be noted, that energy drag discussed here is conceptually different from the E-drag due to direct interlayer energy flux proposed in the context of drag between tightly nested graphene layers \cite{Song}. Magnetothermal drag described by Eq. \eqref{md-rho-th} emerges due to intralayer energy currents. 

\subsection{Magnetodrag in classically strong fields}
 
We find that the initial corrections to drag resistance due to field-dependent viscosity, thermal conductivity, and field-induced shift of the plasmon resonance, persists only at the relatively weak fields. At moderately high fields, but still below the typical scale of plasmon energy, $\omega_c<\omega_{\mathrm{pl}}$, drag magnetoresistivity increases  
\begin{equation}\label{md-rho-mpl}
\rho_D\simeq \rho_Q (TV/\omega^2_{\mathrm{pl}} )(\omega_c/ \omega_{\mathrm{pl}})^2(1/\eta d^2),
\end{equation} 
where $V=e^2/d$ is the scale of Coulomb energy in a confinement of a quantum well. Thus the shape of the full magnetodrag resistance curve is nonmonotonic: the initial negative $H^2$ correction crosses over to a positive $H^2$ growth (we will demonstrate that the positive $\delta \rho^{\mathrm{th}}_D$ correction is in fact smaller than the negative $\delta\rho^{\mathrm{pl}}_{D}$). One should notice that in a striking contrast to the zero-field result [see Eq. \eqref{md-rho-visc}]  Eq. \eqref{md-rho-mpl} scales inversely proportionally to viscosity. For Fermi liquids this implies that $\rho_D\propto T^3 H^2$ (modulo logarithmic factors). The physical mechanism of positive $H^2$ magnetodrag resistance can be traced to the density modulations associated with the viscomagnetic collective modes. In the long wave-length limit, these modes are overdamped and disperse as $q^3$ and $q^4$ due to admixture to optical and acoustic plasmons respectively. Screening is not effective in this region of wave-numbers, and in addition, these modes have a scattering phase-space, which scales as a cube of frequency. As we show below, both factors contribute to strong positive $H^2$ enhancement of drag resistance. At this point, it is perhaps useful to draw an analogy to a related problem where coincidently, similarly dispersing mode, $\omega=\mathrm{i}\varkappa q^3$, exists for composite fermions of the half-filled Landau level \cite{HLR}. From that context it is known that such a density response strongly enhances drag transresistance in quantum Hall bilayers \cite{Ussishkin}. As we alluded, formally the same mechanism of poor screening combined with the phase space argument applies to our situation, albeit for a different microscopic reason and regime of parameters.   
 
\section{Technical approach}\label{sec:technique}
 
\subsection{Stochastic Navier-Stokes equations}
 
We consider a symmetric electronic double-layer system where we assume a steady current in the active layer (layer 1) and no current in the passive layer (layer 2). The equilibrium electron density in each layer is $n$, whereas small density variations and fluctuations of the flow velocity are denoted as $\delta n(\bm{r},t)$ and $\delta \bm{v}(\bm{r},t)$ respectively. For each layer separately we write continuity 
\begin{equation}\label{continuity}
\partial_t n+\dv (n \bm{v})=0,
\end{equation}
and Navier-Stokes equation of motion 
\begin{align}\label{NS}
& m[\partial_t\bm{v}+(\bm{v}\cdot\bm{\nabla})\bm{v}]=\nonumber \\ 
 &-\frac{1}{n}\bm{\nabla}P-\bm{\nabla}U-\frac{e}{c}[\bm{v}\times \bm{H}]+\frac{1}{n}\bm{\nabla}\cdot\bm{\sigma}. 
 \end{align}
The right-hand-side of this equation accounts for forces acting on the liquid from pressure $P(n)$, a self-consistent Coulomb potential $U(n)$, which is determined by the fluid density via the Poisson equation, the Lorentz force, and the tress tensor of viscous fluxes
\begin{equation}\label{stress-tensor}
 \sigma_{ik}=2\eta v_{ik}+(\zeta-\eta)\delta_{ik}\partial_lv_l+\eta_{\mathrm{H}}(\varepsilon_{ijz}v_{kj}+\varepsilon_{kjz}v_{ij})+\varsigma_{ik}.
 \end{equation}
Here $v_{ik}=(\partial_kv_i+\partial_iv_k)/2$, where $i,k$ are Cartesian indices, and we used shorthand notation for the spatial derivative $\partial_i=\partial/\partial x_i$. The notation $\bm{\nabla}\cdot\bm{\sigma}$ in Eq. \eqref{NS} should be understood as a convolution, $\partial_k\sigma_{ik}$, and as usual, summation over the repeated index is assumed here and in all equations below, unless otherwise stated. The tensorial structure of viscous stresses is governed by $v_{ik}$ together with the Kroenecker delta symbol $\delta_{ik}$ and Levi-Civita fully antisymmetric symbol $\varepsilon_{ijk}$. 
Longitudinal stresses are determined by the shear (first) viscosity $\eta$ and bulk (second) viscosity $\zeta$, whereas the transverse stresses occur due to odd (Hall) viscosity $\eta_{\mathrm{H}}$ induced by a field. 
The form of $\sigma_{ik}$ in Eq. \eqref{stress-tensor} requires an explanation. We should recall that in general viscosity is the rank-four tensor $\eta_{ijkl}$. In the external magnetic field, Onsager reciprocal symmetry of kinetic coefficients enforces a condition $\eta_{ijkl}(\bm{H})=\eta_{klij}(-\bm{H})$. As a result, between the unit vector of the field $\bm{e}_H=\bm{H}/H$ and two unity tensors Kronecker ($\delta_{ij}$) and Levi-Civita ($\varepsilon_{ijk}$) there are seven independent tensorial combinations that preserve the symmetry of viscosity tensor so that one needs seven scalar viscosity coefficients \cite{Vol10}. The simplified form of the stress tensor we use in Eq. \eqref{stress-tensor} with three viscosities corresponds to a situation when hydrodynamic flow is constrained to 
the two-dimensional plane and field is oriented perpendicular to that plane. In this formalism, thermal fluctuations are captured by stochastic Langevin forces $\varsigma_{ik}(\bm{r},t)$ 
whose variance is described by the correlation function
\begin{align}\label{Lengevin-stress-corr-fun}
&\langle \varsigma_{ik}(\bm{r},t)\varsigma_{lm}(\bm{r}',t')\rangle=2T\delta(\bm{r}-\bm{r}')\delta(t-t')\nonumber \\
&\times[\eta(\delta_{il}\delta_{km}+\delta_{im}\delta_{kl})+(\zeta-\eta)\delta_{ik}\delta_{lm}],
\end{align}    
where $\langle\ldots\rangle$ denotes thermal averaging. In the classical regime the strength of these fluctuations is obviously determined by the equilibrium temperature $T$. One should notice that even though Hall viscosity enters the stress tensor in Eq. \eqref{stress-tensor} it actually drops out from the correlation function \eqref{Lengevin-stress-corr-fun}. This has to do with the fact that Hall viscosity is dissipationless coefficient as it does not lead to entropy production in the flow and, consequently, does not enter the correlation function by virtue of fluctuation-dissipation relation. In appendix \ref{appendix-1} we show explicitly how terms with $\eta_{\mathrm{H}}$ in $\sigma_{ik}$ cancel out from Eq. \eqref{Lengevin-stress-corr-fun}.   

The steady current $\bm{j}=en\bm{v}$ in the active layer exerts the drag force $\bm{F}_D=\langle \delta n_2(-\bm{\nabla}U_2)$ on the passive layer. Relating the potential to density fluctuations by using the Poisson equation and ignoring the intralayer forces we can express the drag force 
\begin{equation}\label{F-D}
\bm{F}_D=\sum_{q}(-\mathrm{i}\bm{q})\int \mathrm{d}\omega(2\pi e^2/\epsilon q)\mathrm{e}^{-qd}D(q,\omega)
\end{equation}
in terms of the density-density correlation function
 \begin{equation}
D(q,\omega)=\langle\delta n_1(q,\omega)\delta n_2(-q,-\omega)\rangle,
 \end{equation}
 where $\delta n_{1,2}(q,\omega)$ are the Fourier components of the density fluctuations in both layers and $q$ is the absolute value of the vector $\bm{q}$. 
 Knowing the drag force one readily finds the longitudinal magnetodrag resistivity 
 \begin{equation}\label{rho-D-H}
 \rho_D(H,T)=(\bm{v}\cdot\bm{F}_D)/(env)^2.
 \end{equation}
 
 \subsection{Linear response and collective modes}
 
In order to calculate the correlation function of interlayer density fluctuations, and having in mind linear response analysis, we linearize continuity and Navier-Stokes equations with respect to small 
deviations from equilibrium in densities and velocity fluctuations. From Eq. \eqref{continuity} we find for the drive layer 
\begin{equation}\label{continuity-linear}
\partial_t\delta n+\dv (\delta n \bm{v}+n\delta\bm{v})=0. 
\end{equation}
Next, using the explicit form of the stress tensor from Eq. \eqref{stress-tensor} in Eq. \eqref{NS} we find after the linearization 
\begin{align}\label{NS-linear}
\partial_t\delta\bm{v}+(\bm{v}\cdot\bm{\nabla})\delta\bm{v}=-\frac{1}{m}\bm{\nabla}\delta U+\nu_\eta\nabla^2\delta\bm{v}+\nu_\zeta\bm{\nabla}\dv\delta\bm{v}\nonumber \\ 
+\nu_h\{[(\nabla^2\delta\bm{v}+\bm{\nabla}\dv\delta\bm{v})\times\bm{e}_z]+\bm{\nabla}\dv[\delta\bm{v}\times\bm{e}_z]\}/2\nonumber \\ 
-\omega_c[\delta\bm{v}\times\bm{e}_z]+\frac{1}{mn}\bm{\nabla}\cdot\hat{\bm{\varsigma}}
\end{align} 
Here we introduced kinematic viscosities $\nu_\eta=\eta/mn$, $\nu_\zeta=\zeta/mn$, and $\nu_h=\eta_\mathrm{H}/mn$. In the equation above we have neglected terms proportional to gradients of pressure fluctuations $\bm{\nabla}\delta P$ that are driven by both density and entropy fluctuations. These terms contribute to drag resistance, but they are subdominant in the entire temperature range of the collision-dominated regime. We will return to this point later in the text. We have also neglected inductive terms stemming from the induced field fluctuations that render current-current type interlayer interaction. These contributions are small in extra powers of $v/c\ll1$. For the drag layer, linearized equation of motion looks identical to Eq. \eqref{NS-linear}, one only needs to send the drift velocity to zero $\bm v\to0$, but obviously not its fluctuations $\delta\bm{v}(\bm{r},t)$.    

To make further progress in solving these equations it is convenient to make collinear and tangential projections that transform Navier-Stokes equation into a scalar form for two components of the velocity field fluctuations. For this purpose we pass to the Fourier representation $\delta\bm{v}(\bm{r},t)\sim \mathrm{e}^{-\mathrm{i}\omega t+\mathrm{i}\bm{q}\bm{r}}$ and introduce 
\begin{equation}
\delta v_\parallel=(\bm{q}\cdot\delta\bm{v})/q,\quad \delta v_\perp=(\bm{q}\cdot[\delta\bm{v}\times\bm{e}_z])/q.
\end{equation} 
Then separating Eq. \eqref{NS-linear} for both velocity components we find 
\begin{align}
&[-\mathrm{i}\omega+\mathrm{i}(\bm{v}\cdot\bm{q})]\delta v_\parallel=\nonumber \\
&-\frac{\mathrm{i}q}{m}\delta U-\omega_\nu\delta v_\parallel-(\omega_c+\omega_h)\delta v_\perp+\frac{\mathrm{i}}{mnq}(\bm{q}\cdot\hat{\bm{\varsigma}}\bm{q}),\\
&[-\mathrm{i}\omega+\mathrm{i}(\bm{v}\cdot\bm{q})]\delta v_\perp=\nonumber \\
&-\omega_\eta\delta v_\perp+(\omega_c+\omega_h)\delta v_\parallel+\frac{\mathrm{i}}{mnq}(\bm{q}\cdot[\hat{\bm{\varsigma}}\bm{q}\times\bm{e}_z]),
\end{align}
where $\omega_\eta=\nu_\eta q^2$, $\omega_\nu=\nu q^2$, $\omega_h=\nu_\mathrm{H}q^2$, and $\nu=\nu_\eta+\nu_\zeta$. It is worth noting that $\nu_\mathrm{H}<0$ for $\omega_c>0$, so the $|\nu_\mathrm{H}| q^2$ term is actually subtracted from $\omega_c$. For small density fluctuations, $\delta n/n\ll 1$, the linear screening approximation applies and the variations of the Coulomb potential in the active layer $\delta U_1$ can be expressed via the variations of electron density in the both layers
 as follows 
 \begin{equation}
\delta U_1=\frac{2\pi e^2}{\epsilon q}(\delta n_1+\mathrm{e}^{-qd}\delta n_2). 
 \end{equation}
For the passive layer this expression is the same, one only needs to flip indices $1\leftrightarrow2$. We can use continuity equation \eqref{continuity-linear} to exclude $\delta v_\parallel$ and $\delta v_\perp$, and obtain a closed set of equations for density fluctuations only in both layers $\delta n_{1,2}$. Motivated by the physical picture of propagating in-phase and out-of-phase density modes it will be convenient to introduce symmetric and antisymmetric combinations $\delta n_\pm=\delta n_1\pm\delta n_2$, including the Langevin fluxes $\hat{\varsigma}_\pm=\hat{\varsigma}_1\pm\hat{\varsigma}_2$. These steps enable us to write density response in a combined form 
\begin{align}\label{Pi-delta-n}
&\Pi_{\pm}\delta n_\pm=(\omega_c+\omega_h)\frac{q^2\varsigma^\perp_\pm}{m}-(\omega_\eta-\mathrm{i}\omega)\frac{q^2\varsigma^\parallel_\pm}{m} \nonumber \\ 
&+\frac{\mathrm{i}(\bm{v}\cdot\bm{q})}{2}\left[\Gamma_+\delta n_++\Gamma_-\delta n_-
-\frac{q^2(\varsigma^\parallel_++\varsigma^\parallel_-)}{m}\right].
\end{align} 
Here the projections of Langevin fluxes are 
\begin{equation}
\varsigma^\perp_\pm=\frac{\bm{q}\cdot[\hat{\bm{\varsigma}}_\pm\bm{q}\times\bm{e}_z]}{q^2},\quad 
\varsigma^\parallel_\pm=\frac{\bm{q}\cdot(\hat{\bm{\varsigma}}_\pm\bm{q})}{q^2},
\end{equation}
and the polarization functions have the form 
\begin{align}\label{Pi}
\Pi_\pm(q,\omega)=\mathrm{i}\omega(\omega_\eta-\mathrm{i}\omega)(\omega_\nu-\mathrm{i}\omega)-(\omega_\eta-\mathrm{i}\omega)\omega^2_\pm \nonumber \\ 
+\mathrm{i}\omega(\omega_c+\omega_h)^2.
\end{align} 
The dynamic vertex functions that couple fluctuating density modes in the presence of the drift and finite magnetic field are 
\begin{align}
\Gamma_\pm(q,\omega)=(\omega_\eta-\mathrm{i}\omega)(\omega_\nu-\mathrm{i}\omega)-\mathrm{i}\omega(\omega_\eta+\omega_\nu)\nonumber \\
+\omega^2_\pm+(\omega_c+\omega_h)^2.
\end{align} 
Here we introduced symmetric/antisymmetric plasmon frequencies $\omega^2_\pm=\omega^2_p(1\pm\mathrm{e}^{-qd})$ and $\omega^2_p=2\pi e^2nq/m\epsilon$. 

\begin{figure}
  \centering
  \includegraphics[width=3.25in]{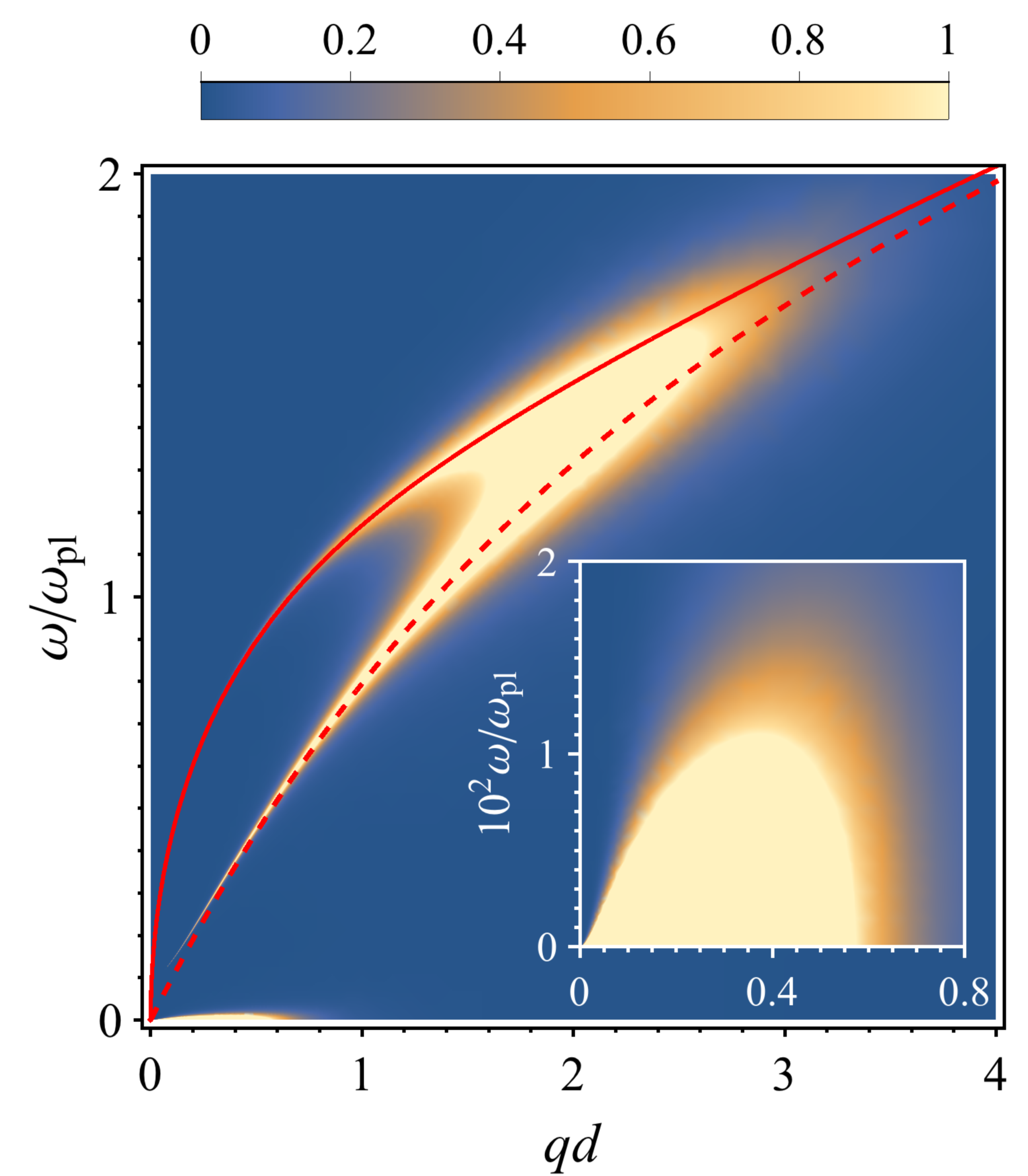}
  \caption{A contrast plot for the density-density correlation function $D(q,\omega)$ displaying in a.u. the intensity of hydrodynamic collective modes in a bilayer of 2D electron liquids. Plasmon branches are highlighted by the solid line for optical mode, and dashed line for acoustic mode, respectively. Inset plot shows a zoom into the low-momentum region of viscomagnetic modes.}  
  \label{Fig-modes}
\end{figure}

It is instructive to investigate properties of the polarization function more closely in few limiting cases. Ignoring viscous effects entirely and without magnetic field Eq. \eqref{Pi} simplifies to $\Pi_\pm=\mathrm{i}\omega(\omega^2_\pm-\omega^2)$, which has poles at the real axis of frequency corresponding to plasmon resonances. This is obviously physically expected as in general poles of the polarization function correspond to collective modes propagating in the system. At finite field, but still without viscosity terms, poles simply shift to higher energies, $\Pi_\pm=\mathrm{i}\omega(\omega^2_\pm+\omega^2_c-\omega^2)$, as expected for magnetoplasmon modes.  For weak viscous effects, $\omega_{\eta,\nu}\ll\omega_\pm$, poles shift from the real axis into the complex plane of frequency, so that plasmons acquire a finite life time $\tau^{-1}_{\mathrm{pl}}=(\omega_\eta/2)[\omega_\nu/\omega_\eta+\omega^2_c/(\omega^2_\pm+\omega^2_c)]$. This energy scale sets the width of plasmon resonance. Odd viscosity also modifies real part of the magnetoplasmon dispersion curve by inducing a shift by $\Delta\omega_\pm(H,q)=\omega_c\omega_h/(\omega^2_\pm+\omega^2_c)^{1/2}$. Importantly, the viscous pole couples to plasmons in magnetic field, $\mathrm{i}\omega\to\mathrm{i}\omega+\omega_\eta \omega^2_{\pm}/(\omega^2_\pm+\omega^2_c)$, that gives rise to overdamped viscomagnetic collective modes. In particular, for the admixture of the viscous diffusive pole to the optical plasmon branch we have in the limit of long wave-length fluctuations $\omega=\mathrm{i}\varkappa q^3$, where $\varkappa=2\pi e^2\eta/(\epsilon m^2\omega^2_c)$. This discussion is further exemplified by Fig. \ref{Fig-modes} where we plot the density response function $D(q,\omega)$ of emergent hydrodynamic modes. 

We seek for the solution of Eq. \eqref{Pi-delta-n} to the linear order in $\bm{v}$ in the form $\delta n_\pm=\delta n^{(0)}_\pm+\delta n^{(1)}_\pm$, where   
\begin{align}\label{delta-n-0}
&\delta n^{(0)}_\pm=\frac{q^2}{m\Pi_\pm}\left[(\omega_c+\omega_h)\varsigma^\perp_\pm-(\omega_\eta-\mathrm{i}\omega)\varsigma^\parallel_\pm\right], \\ 
&\delta n^{(1)}_\pm=\frac{\mathrm{i}(\bm{v}\cdot\bm{q})}{2\Pi_\pm}\left[\Gamma_+\delta n^{(0)}_++\Gamma_-\delta n^{(0)}_-
-\frac{q^2(\varsigma^\parallel_++\varsigma^\perp_-)}{m}\right].
\end{align}
These expressions allow us to calculate the density-density correlation function 
\begin{align}\label{P}
&D(q,\omega)=\frac{\mathrm{i}(\bm{q}\cdot\bm{v})(nTq^2/m)}{|\Pi_+|^2|\Pi_-|^2} \sum_\pm\pm\Pi^*_{\mp} \nonumber \\ 
&\left[\Gamma_\pm(\omega_\eta(\omega_c+\omega_h)^2+\omega_\nu(\omega^2+\omega^2_\eta))+\Pi_\pm\omega_\nu(\omega_\eta+\mathrm{i}\omega)\right],
\end{align}
where we made use of the following thermal averages 
\begin{align}
\langle \varsigma^\parallel_i\varsigma^\parallel_j\rangle
 =4T(\eta+\zeta)\delta_{ij}, \quad \langle\varsigma^\parallel_i\varsigma^\perp_j\rangle
 =0, \quad \langle \varsigma^\perp_i\varsigma^\perp_j\rangle
 =4T\eta\delta_{ij},
\end{align}
that were obtained with the help of Eq. \eqref{Lengevin-stress-corr-fun} and indices $\{i,j\}=\pm$ imply symmetric/antisymmetric combinations. 

In order to find resulting drag resistance we need to use $D(q,\omega)$ in Eqs. \eqref{F-D} and \eqref{rho-D-H}. A complete functional form of the field dependence in $\rho_D(H)$ is rather complex and difficult to find analytically. Instead, we will inspect Eq. \eqref{P} in several formal limiting cases that enables us to uncover more clearly physical processes responsible for resistance. We are also able to find an analytic interpolation formula that captures temperature and field dependence of drag resistance in the parameter space of interest. A complete solution can be obtained rather efficiently by numerical integration. We will discuss below a convenient choice of dimensionless variables for numerical analysis.

\subsection{Viscous flow magnetodrag resistance} 

The dependence of the drag resistance on the magnetic field stems from several main sources. The first one is just due to the field dependence of the viscosity itself [see Eq. \eqref{eta-xx}].   
In order to single out this contribution from the density response function in Eq. \eqref{P} one needs to send terms with $\omega_c+\omega_h$ to zero in the polarization function $\Pi_\pm$ and 
vertex function $\Gamma_\pm$, but retain field dependence of $\omega_\eta$ and $\omega_\nu$. The corresponding part of the density-density correlation function then reads 
\begin{align}\label{P-visc}
D_{\mathrm{visc}}=\frac{\mathrm{i}(\bm{v}\cdot\bm{q})(nTq^2/m)(\omega^2_+-\omega^2_-)\omega^2_\nu}
{[\omega^2\omega^2_\nu+(\omega^2-\omega^2_-)^2][\omega^2\omega^2_\nu+(\omega^2-\omega^2_+)^2]}
\end{align}
Inserting this expression into Eqs. \eqref{F-D} and \eqref{rho-D-H}, frequency integration can be readily done by poles in the complex plane of $\omega$, and remaining 
momentum integration can be made dimensionless by introducing a variable $x=qd$. We thus find 
\begin{equation}\label{rho-D-0}
\rho^{\mathrm{visc}}_D(H)=\frac{\epsilon T(\eta+\zeta)}{16\pi^2 e^4n^4d^5}F_0(\beta)
\end{equation}
where we introduced dimensionless parameter 
\begin{equation}\label{beta}
\beta=\frac{\eta+\zeta}{n}\sqrt{\frac{a_B}{2\pi nd^3}},
\end{equation}
and a dimensionless function of that parameter 
\begin{equation}\label{F-0}
F_0=\int\limits^{\infty}_{0}\frac{x^4(\beta^2 x^3+1)\mathrm{e}^{-x}\mathrm{d}x}{2\sinh(x)(\beta^2 x^3+\mathrm{e}^{-2x})}.
\end{equation}
Provided that $k_Fd>1$, which is our basic assumption for the weakly coupled layers, it will be shown below that $\beta<1$ in the high temperature hydrodynamic regime. In the limit $\beta\to0$, the dimensionless function $F_0(\beta)$ diverges logarithmically (see appendix \ref{appendix-2} for a detailed analysis). Retaining the leading asymptotic behavior and expanding Eq. \eqref{rho-D-0} to $H^2$ order we obtain magnetic field induced correction to viscous Coulomb drag resistance 
\begin{equation}
\delta\rho^{\mathrm{visc}}_D(H)\simeq -\rho_Q\frac{(\eta/n)(\omega_c\tau_{\mathrm{ee}})^2}{r_s(k_Fd)^5}\frac{T}{E_F}\ln^5\left[\frac{1}{\beta^2\ln^3(1/\beta)}\right]. 
\end{equation}  
In this expression (and in what follows) we neglected bulk viscosity, as it is typically small compared to the sheer viscosity. 
This is certainly justified for Fermi liquids where $\zeta\sim (T/E_F)^2\eta$ so that $\zeta\ll\eta$ at all temperatures below the Fermi energy $T\ll E_F$. We have also suppressed all numerical factors for brevity and kept only parametric and functional dependence on physical parameters. 

\begin{figure}
  \centering
  \includegraphics[width=3.25in]{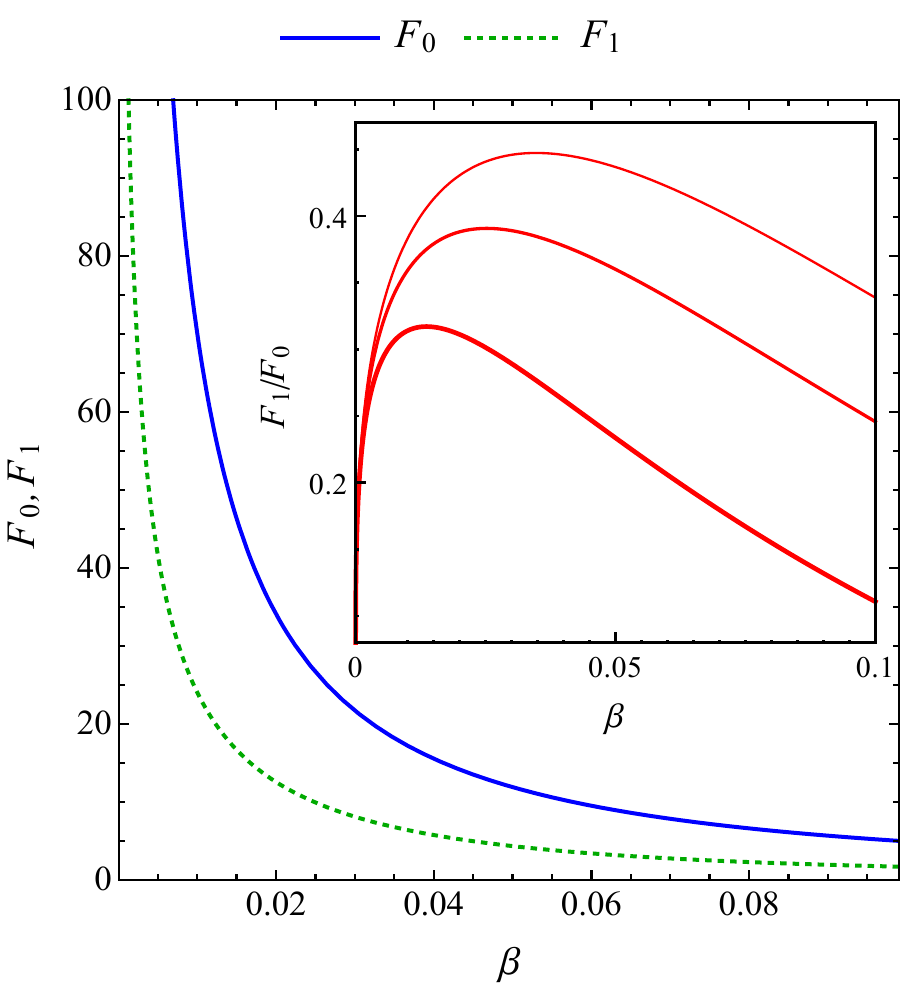}
  \caption{(color online) Dependence of the dimensionless functions $F_{0,1}$ from Eqs. \eqref{F-0} and \eqref{F-1} on viscosity parameter $\beta$. 
  In the main panel we plotted $F_0(\beta)$ and $F_1(\alpha =\beta,\beta)$ in solid and dashed lines, respectively. 
  In the inset we show the ratio $F_1(\alpha=z\beta,\beta)/F_0(\beta)$ for several values of $z=1/4, 1/2, 1$ displayed in thin, normal, thick lines, respectively.}\label{Fig-F}
\end{figure}

\subsection{Magnetoplasmon drag resistance}

The second source of drag resistance captured by Eq. \eqref{P} is induced by the coupling between the longitudinal and transverse modes of fluctuations facilitated by the magnetic field. 
As a result, both electron cyclotron motion and Hall viscosity terms modify plasmon dispersion so that this mechanism of drag is strongest at the plasmon resonance of a bilayer system. 
Expanding Eq. \eqref{P} to the second order in $\omega_c+\omega_h$ 
we get $D=D_{\mathrm{visc}}+D_{\mathrm{mpl}}$ where 
\begin{align}\label{P-mpl}
&D_{\mathrm{mpl}}=\frac{-2\mathrm{i}(\bm{v}\cdot\bm{q})(nTq^2/m)(\omega^2_+-\omega^2_-)\omega^2_\eta}
{[\omega^2\omega^2_\eta+(\omega^2-\omega^2_-)^2]^2[\omega^2\omega^2_\eta+(\omega^2-\omega^2_+)^2]^2} \nonumber \\ 
&\times\frac{(\omega_c+\omega_h)^2}{(\omega^2+\omega^2_\eta)}\left[\omega^2\omega^2_\eta+(\omega^2-\omega^2_+)(\omega^2-\omega^2_-)\right] \nonumber \\ 
&\times\left[\omega^2\omega^2_\eta-(\omega^2-\omega^2_+)(\omega^2-\omega^2_-)+\omega^2(\omega^2_++\omega^2_--2\omega^2)\right].
\end{align} 
This cumbersome expression can be split into the sum of simple fractions each of which has relatively simple pole structure that can be then integrated over the frequency analytically. As in the previous section, the final momentum integration can be made dimensionless. We thus find from Eqs. \eqref{P-mpl}, \eqref{F-D} and \eqref{rho-D-H} the corresponding resistance 
\begin{equation}\label{rho-D-mpl}
\delta\rho^{\mathrm{mpl}}_D(H)=-\frac{\epsilon T\eta(\omega_c/\omega_{\mathrm{pl}})^2}{16\pi^2e^4n^4d^4}F_1(\alpha,\beta)
\end{equation}
where 
\begin{equation}\label{F-1}
F_1=\int\limits^{\infty}_{0}\frac{3\beta^2x^6(1-2\alpha x^2)\mathrm{e}^{-x}\mathrm{d}x}{2\sinh(x)[\beta^2x^3+\mathrm{e}^{-2x}]^2[(1+2\beta^2x^3)^2-\mathrm{e}^{-2x}]}, 
\end{equation}
and $\alpha=R_c|\nu_\mathrm{H}|/(v_Fd^2)$ with $R_c=v_F/\omega_c$ being the cyclotron radius. In $F_1(\alpha,\beta)$ we kept only the linear in $\alpha$ term and neglected all higher order contributions. 
We have also retained only leading order terms in powers of $\beta$ in the numerator of the integrand. The product of $R_c|\nu_\mathrm{H}|$ is field independent up to correction of the order of $\omega^2_c$, which are of higher order of accuracy for Eq. \eqref{rho-D-mpl} since it is already quadratic in $H$. 

To make further connection to our earlier qualitative discussions, we can formally split Eq. \eqref{rho-D-mpl} into two terms $\delta\rho^{\mathrm{mpl}}_D=\delta\rho^{\mathrm{pl}}_{D}+\delta\rho^{\mathrm{H}}_{D}$.  The first resonant magnetoplasmon contribution follows from the part of $F_1$ that is independent of $\alpha$. Retaining the leading asymptotic behavior of $F_1$ when $\beta\ll1$ (see appendix \ref{appendix-2} for details) we find 
\begin{equation}
\delta\rho^{\mathrm{pl}}_{D}\simeq-\rho_Q\frac{(\eta/n) (\omega_c/\omega_{\mathrm{pl}})^2}{r_s(k_Fd)^5}\frac{T}{E_F}\ln^3\left[\frac{1}{\beta^2\ln^3(1/\beta)}\right].
\end{equation}       
The second linear in $\alpha$ term of $F_1$ defines the Hall viscosity contribution to magnetoplasmon drag resistance 
\begin{equation}
\delta\rho^{\mathrm{H}}_{D}\simeq-\rho_Q\frac{\eta\eta_{\mathrm{H}} (k_FR_c) (\omega_c/\omega_{\mathrm{pl}})^2}{r_sn^2(k_Fd)^7}\frac{T}{E_F}\ln^5\left[\frac{1}{\beta^2\ln^3(1/\beta)}\right].
\end{equation}  
On Fig. \ref{Fig-F} we plot $F_{0,1}$ functions as well as their ratio that defines the relative contribution to magnetodrag resistance. 

\subsection{Thermomagnetic drag resistance}

In order to capture the thermal component of the drag resistance $\delta\rho^{\mathrm{th}}_{D}$ quoted in Eq. \eqref{md-rho-th} we need to invoke an extended hydrodynamic theory.  This requires an account of extra forces $\partial_i P$ from the pressure term in the Navier-Stokes equation. Even though the equation of state may not be known for a strongly correlated liquid, still one can relate pressure fluctuations $\delta P=(\partial P/\partial n)_S\delta n+(\partial P/\partial s)_V\delta s$ in a given layer to density ($\delta n$) and entropy $(\delta s)$ fluctuations via general thermodynamic relations such as $(\partial P/\partial s)_V=n^2(\partial  T/\partial n)_S$. Due to screening the compressibility term in pressure fluctuations can be neglected in the long wavelength limit, however the entropic term must be retained. To close the system of equations we then need an entropy production equation whose linearized version reads in Fourier components 
\begin{equation}
nT[\mathrm{i}(\bm{q}\cdot\bm{v})-\mathrm{i}\omega]\delta s=-\kappa q^2\delta T-\mathrm{i}(\bm{q}\cdot\bm{g})
\end{equation}     
where $\kappa$ is the thermal conductivity of electron liquid and $\bm{g}$ is the Langevin force of thermal fluxes  that is described by a familiar correlation function [see appendix \eqref{appendix-1} for a reminder]
\begin{equation}\label{Langevin-thermal-corr-fun}
\langle g_i(\bm{r},t)g_j(\bm{r}',t')\rangle=2\kappa T^2\delta_{ij}\delta(\bm{r}-\bm{r}')\delta(t-t'). 
\end{equation} 
For density and entropy as being independent thermodynamic variables, temperature fluctuations $\delta T=(\partial T/\partial n)_S\delta n+(\partial T/\partial s)_V\delta s$ can be related to isentropic expansion of the liquid $\alpha_S=-(\partial \ln n/\partial T)_S$ and specific heat $c_V=T(\partial s/\partial T)_V$. Thermal fluxes in the fluid give raise to diffusively propagating overdamped mode of entropy fluctuations $\omega=\mathrm{i}\kappa q^2/nc_V$ and propagating acoustic mode $\omega_\alpha=u_T q$, with the characteristic velocity $u_T=\sqrt{T/mc_V}(\partial s/\partial\ln n)_T$, of thermal expansion that lead to an additional density variation 
\begin{equation}
\delta n^{\mathrm{th}}_\pm=-\frac{\mathrm{i}u_Tq^2}{\sqrt{mc_VT}}\frac{(\bm{q}\cdot\bm{g}_\pm)}{\Pi_\pm}
\end{equation}
that should be added to $\delta n^{(0)}_\pm$ in Eq. \eqref{delta-n-0}. Polarization function $\Pi_\pm$ is also modified by a term $-(\kappa q^2/nc_V)\omega^2_\alpha$ that should be included in Eq. \eqref{Pi}. 
As shown in Ref. \cite{Apostolov} these modes couple to plasmons and render an additional drag resistance 
\begin{equation}
\delta\rho^{\mathrm{th}}_{D}=\frac{3\zeta(3)}{32\pi e^2}\frac{1}{nd^2}\frac{\epsilon^2T}{\chi n}\left(\frac{u_T}{\omega_{\mathrm{pl}}d}\right)^4
\end{equation} 
where $\chi=\kappa/n c_V$ is the kinematic thermal diffusivity. At finite magnetic field we need to account for the modification of thermal conductivity that gets a correction $\delta \kappa=-\kappa (\omega_c\tau_{\mathrm{ee}})^2$ which then translates into Eq. \eqref{md-rho-th}. The coupling of the thermal mode to magnetoplasmon gives an additional contribution to drag resistance that, however, has an extra smallnesses in powers $(u_T/\omega_{\mathrm{pl}}d)$, and thus can be neglected. 

\subsection{Viscomagnetic drag resistance}

\begin{figure}
  \centering
  \includegraphics[width=3in]{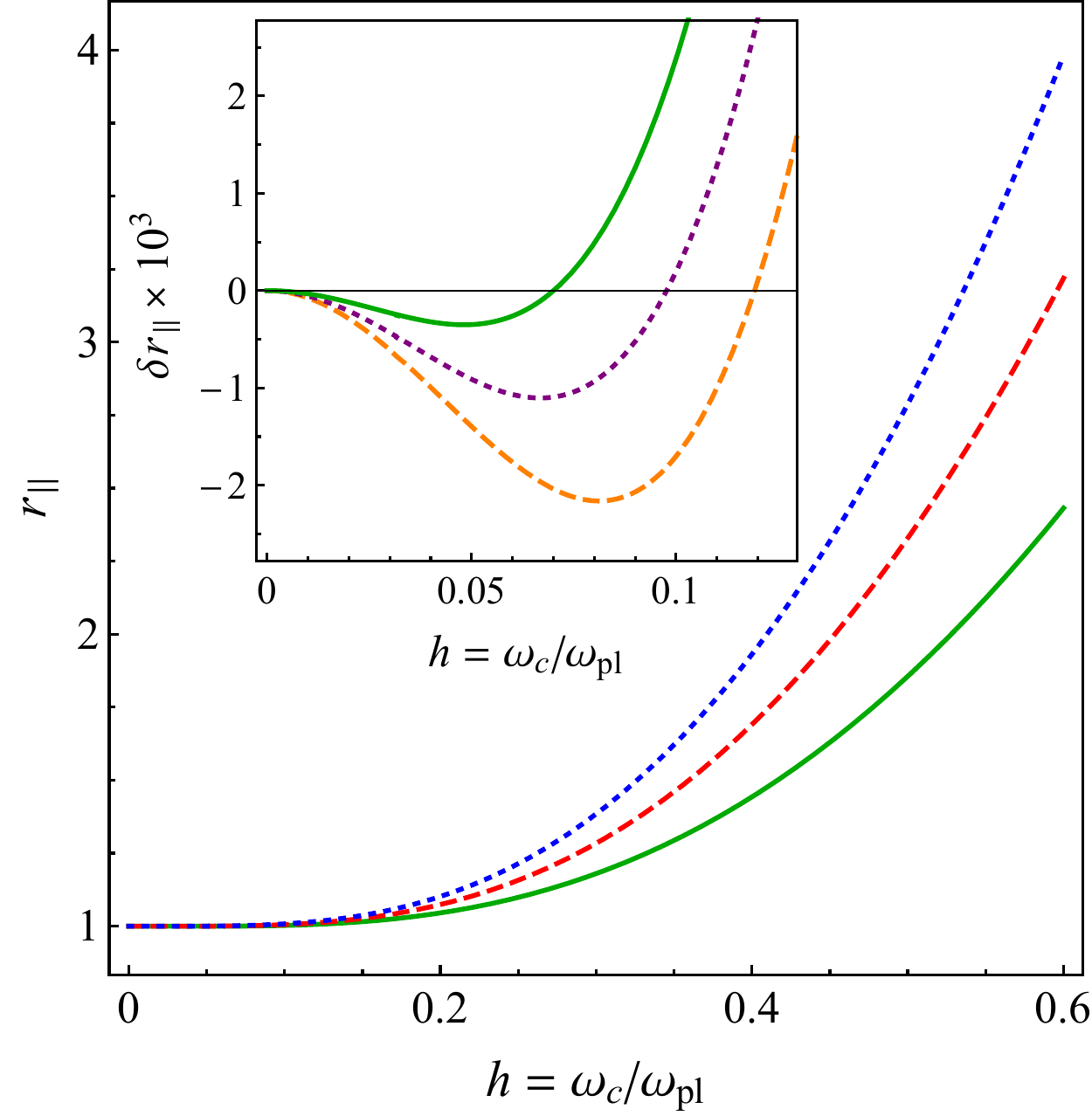}
  \caption{In the main panel we show magnetodrag resistance normalized to its zero-field value, $r_\parallel=\rho_D(H,T)/\rho_D(0,T)$, plotted for several different values of the viscosity parameter $\beta=1/8, 1/14, 1/20$ [Eq. \eqref{beta}] as a function of cyclotron frequency normalized to the energy scale of a plasmon $h=\omega_c/\omega_{\mathrm{pl}}$. The inset plot shows a zoom into the low-field region where $\delta r_\parallel=r_\parallel-1$ is plotted in dashed, dotted, solid lines corresponding to different values of a ratio $\alpha/\beta=1/4, 1/2, 1$ and fixed $\beta=1/8$. } 
  \label{Fig-drag}
\end{figure}

A direct expansion of $D(q,\omega)$ at small fields, that we carried out in the previous subsections, captures correctly only the initial trend of magnetoresistance. Our carefully numerical analysis revealed that at higher fields, drag in fact growths as a function of $H$. To develop an analytical understanding of apparent nonmonotonic field dependence, we single out poles in the polarization functions $\Pi_\pm(q,\omega)$ corresponding to all the collective modes in the system. Specifically, Eq. \eqref{Pi} can be written as follows
\begin{align}\label{Pi-approx}
&\Pi_\pm(q,\omega)=(\mathrm{i}\omega-\lambda_\pm\omega_\eta)\nonumber \\ 
&[(\mathrm{i}\omega)^2+\omega^2_\pm/\lambda_\pm-\mathrm{i}\omega[\omega_\nu+(1-\lambda_\pm)\omega_\eta]],
\end{align}         
where $\lambda_\pm$ is the root of the following algebraic equation
\begin{equation}
\omega^2_\pm/\lambda_\pm=(\omega_c+\omega_h)^2+\omega^2_\pm+\omega_\eta(\omega_\nu-\lambda_\pm\omega_\eta)(1-\lambda_\pm).
\end{equation}
In the limit of long wave-length fluctuations $\omega_\eta\sim\omega_\nu\ll\omega_\pm$, so that this equation can be easily solved iteratively in small viscous terms. 
To the main order in $O(\omega^2_\eta)$ corrections 
\begin{equation}
\lambda_\pm=\omega^2_\pm/[(\omega_c+\omega_h)^2+\omega^2_\pm]. 
\end{equation}
With this form of $\Pi_\pm$ in Eq. \eqref{Pi-approx} we can perform frequency integral analytically in the expression for the drag force. We find 
\begin{align}
&\int\limits^{+\infty}_{-\infty}D(q,\omega)\frac{\mathrm{d}\omega}{2\pi}=\mathrm{i}(\bm{v}\cdot\bm{q})\frac{nTq^2}{m}\nonumber \\  
&\frac{\mathrm{e}^{-2qd}[\omega^2_\eta(\omega^6_p+\omega^2_H\omega^4_p+4\omega^6_H)+\omega^4_p\omega^4_c\mathrm{e}^{-2qd}]}{2\omega^4_p\omega_\eta[\omega^2_\eta(\omega^2_p+2\omega^2_H)^2+\omega^4_p(\omega^2_p+\omega^2_H)\mathrm{e}^{-2dq}]}
\end{align} 
where $\omega_H=\omega_c+\omega_h$. To arrive at this expression, we have also made a consistent approximation by treating $\omega_\eta\sim (\omega^2_+-\omega^2_-)/\omega_p\ll \omega_p$. As a result, for the magnetodrag resistance due to resonant collective modes we obtain the interpolation formula 
\begin{equation}
\rho_D(H,T)=\frac{\epsilon T\eta}{16\pi^2e^4n^4d^5}F_3(\alpha,\beta,h), \quad h=\omega_c/\omega_{\mathrm{pl}}, 
\end{equation}
where 
\begin{equation}
F_3\!=\!\!\int\limits^{\infty}_{0}\!\frac{[\beta^2x^2(4h^6(x)\!+\!h^2(x)x^2\!+\!x^3)\!+\!h^4(x)\mathrm{e}^{-2x}]\mathrm{e}^{-2x}\mathrm{d}x}{\beta^2[\beta^2x^2(2h^2(x)+x)^2+(h^2(x)+x)\mathrm{e}^{-2x}]}
\end{equation}
and $h(x)=h(1-\alpha x^2)$. It should be noted that integrand in $F_3$ has two extremal regions as function of $x$. The first one is at $x\ll1$, due to viscomagnetic collective mode, and the second at $x\sim \ln(1/\beta)\gg1$, due to plasmon resonances. Function $F_3$ describes a smooth crossover from the negative quadratic drag resistance at weak fields [Eq. \eqref{rho-D-mpl}] to positive quadratic magnetodrag at a higher field region [Eq. \eqref{md-rho-mpl}]. The shape of $F_3$ is illustrated in Fig. \ref{Fig-drag} for several different parameters. 

 \section{Analysis and discussion}\label{sec:discussion}

Results obtained in this work for magnetodrag resistance cover a broad range of temperatures, electron densities, and interaction strengths. A particular temperature and field dependence of $\rho_D(H,T)$ and relative magnitude of various terms depends on microscopic state of electronic fluid via viscosities $\eta(T,H), \zeta(T,H)$, and thermal conductivity $\kappa(T,H)$ coefficients, as well thermodynamic characteristics such as specific heat $c_V(T,H)$. For strongly correlated liquids, a detailed microscopic theory of the temperature dependence of all these coefficients has not been developed. To narrow down the parameter space of different transport regimes in our comparative analysis we choose to focus on Fermi liquids at $r_s\sim 1$ where $\kappa\simeq E^2_F/(T\ln(E_F/T))$, $\eta\simeq n(E_F/T)^2\ln^2(E_F/T)$, $\tau^{-1}_{\mathrm{ee}}\simeq (T^2/E_F)\ln(E_F/T)$ and $c_V\simeq T/E_F$ \cite{Abrikosov,Brooker-Sykes,Novikov,Mishchenko,Chubukov-Maslov}. For this case, the parameter $\beta$ in Eq.\eqref{beta} that defines logarithmic renormalization of drag resistance from plasmon resonances is given by $\beta\simeq1/(k_Fd)^{3/2}(E_F/T)^2$.  
For brevity, we shall further omit these extra logarithmic factors from $F$-functions of Eqs. \eqref{F-0} and \eqref{F-1}, including also FL logarithms.

We would like to note that the discussion below can be generalized to the non-degenerate limit, $T>E_F$, whereby electrons form a classical gas, so that $\eta\sim mv_TT/e^2$ and $\kappa\sim v_TT/e^2$, where $v_T\sim\sqrt{T/m}$ is the thermal velocity. Another interesting regime is the so-called   semi-quantum liquid, for which some conjectures were put forward in Refs. \cite{Andreev-Kosevich,Kivelson}. In particular, $c_V\propto T$, $\kappa\propto T$, $\eta\propto 1/T$ at zero magnetic field.

\subsection{Temperature dependence}

\begin{figure}
  \centering
  \includegraphics[width=3.5in]{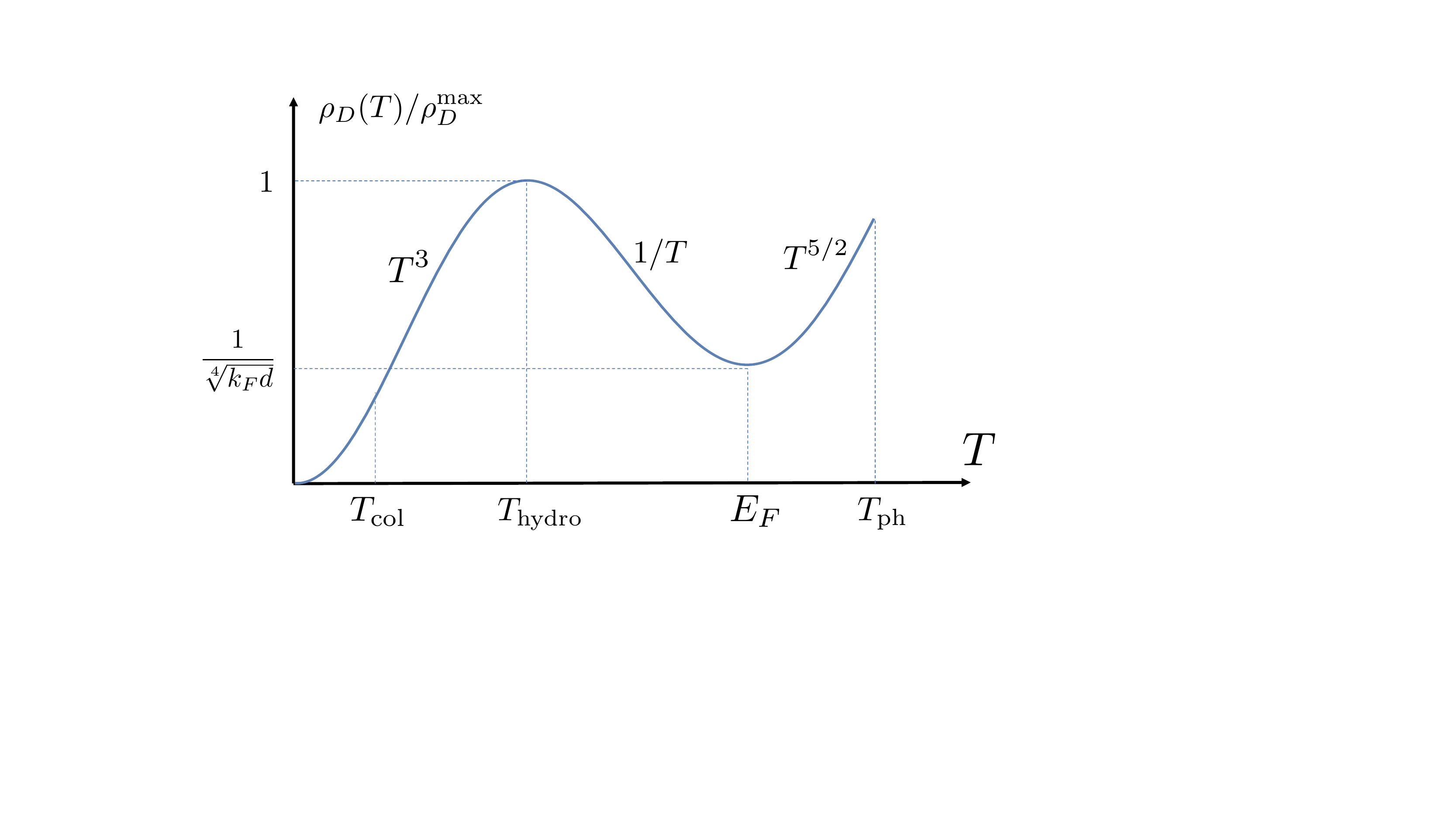}
  \caption{A schematic plot for the temperature dependence of the zero-field drag resistance normalized to its maximal value at the plasmon peak. The range of temperatures $T_{\mathrm{col}}<T<T_{\mathrm{hydro}}$ describes the crossover from collisionless to hydrodynamic regime \cite{Chen}. Hydrodynamic regime applies 
  at higher temperatures, $T>T_{\mathrm{hydro}}$, until its is destroyed by electron momentum and energy non-conserving collisions with phonons at $T\sim T_{\mathrm{ph}}$. Power-exponents in a crossover $T^3\to 1/T$ are quoted assuming FL picture.} 
  \label{Fig-r-D-T}
\end{figure}

The zero-field longitudinal drag resistance has a maximum at a temperature scale $T_{\mathrm{pl}}\sim E_F/\sqrt[4]{k_Fd}$, where it can be estimated to be of the order of $\rho^{\mathrm{max}}_D/\rho_Q\sim1/(k_Fd)^{19/4}$ \cite{Apostolov}. The scale of $T_{\mathrm{pl}}\sim T_{\mathrm{hydro}}$ marks the condition at which plasmons attain hydrodynamic limit $\omega_{\mathrm{pl}}\sim\tau^{-1}_{\mathrm{ee}}(T_{\mathrm{pl}})$. 
It is important to stress that a simple extrapolation of the drag resistance from collisionless side $T<T_{\mathrm{col}}$ (namely a contribution of the particle-hole continuum) underestimates drag resistance in a parametrically large factor. Indeed, as particle-hole excitations give quadratic in temperature drag $\rho^{\mathrm{p-h}}_D/\rho_Q \sim 1/(k_Fd)^4(T/E_F)^2$ at lowest temperatures \cite{Kamenev}, this behavior changes to linear  $\rho^{\mathrm{p-h}}_D/\rho_Q \sim 1/(k_Fd)^5(T/E_F)$ at $T\sim E_F/(k_Fd)$ due to kinematic constraints. Extrapolating then this $\rho^{\mathrm{th}}_D$ to $T_{\mathrm{pl}}$ we see that $\rho^{\mathrm{p-h}}_D/\rho^{\mathrm{max}}_D\sim 1/\sqrt{k_Fd}$. The proper crossover in the region $T_{\mathrm{col}}<T<T_{\mathrm{hydro}}$ is captured by a low-energy tail of plasmons which is beyond pure hydrodynamic theory and requires solution of the full kinetic equation \cite{Chen}. The important conclusion that can be drawn from this naive extrapolation attempt is that intralayer equilibration processes strongly enhance drag and this is qualitatively consistent with experimental findings (see for example Ref. \cite{Hill-2} where a detailed discussion is presented about the discrepancies of observation and theory predictions from improved random-phase-approximation). At temperatures above $T_{\mathrm{pl}}$ the zero-field drag resistance decays from its maximal value as $\rho_D\simeq\rho_Q/(k_Fd)^5(E_F/T)$, at $T_{\mathrm{hydro}}<T<E_F$, 
however, it starts to grow again in the nondegenerate regime at $T>E_F$ as
$\rho_D\simeq\rho_Q/(k_Fd)^5(T/E_F)^{5/2}$, which persists until the hydrodynamic description breaks down by frequent electron-phonon scattering processes [see Fig. \ref{Fig-r-D-T} for the illustration].  

\subsection{Magnetic field dependence}

Further, we have investigated how maximum of drag is affected by magnetic field. At lowest fields, considering the ratio between plasmon and viscous contributions to magnetodrag resistance from Eqs. \eqref{md-rho-visc} and \eqref{md-rho-pl} 
\begin{equation}
\frac{\delta\rho^{\mathrm{pl}}_{D}}{\delta\rho^{\mathrm{visc}}_D}\sim\frac{1}{(\omega_{\mathrm{pl}}\tau_{\mathrm{ee}})^2}\sim (k_Fd)(T/E_F)^4,
\end{equation} 
we see that at the onset of hydrodynamic regime, $T\sim T_{\mathrm{pl}}\sim T_{\mathrm{hydro}}$, these contributions are of the same order, however plasmon mechanism dominates, $\delta\rho^{\mathrm{pl}}_{D}>\delta\rho^{\mathrm{visc}}_D$, everywhere at higher temperatures $T_{\mathrm{pl}}<T<E_F$. The Hall viscosity contribution contains an extra factor in square of the Knudsen number. At the plasmon resonance, $T\sim T_{\mathrm{pl}}$, we can estimate $\mathrm{Kn}\sim1/\sqrt{k_Fd}$, so that $\delta\rho^{\mathrm{H}}_D<\delta\rho^{\mathrm{pl}}_D$. The thermomagnetic part of drag resistance is a growing function of temperature but even at the plasmon resonance it is still small compared to plasmonic contribution, $\delta\rho^{\mathrm{th}}_{D}/\delta\rho^{\mathrm{pl}}_{D}\sim 1/(k_Fd)$, and becomes of order one, $\delta\rho^{\mathrm{th}}_{D}/\delta\rho^{\mathrm{pl}}_{D}\sim 1$, only at $T\sim E_F$. In summary, for a FL regime the temperature scalings of these terms are as follows (again omitting both FL logarithms and plasmon resonance logarithms): $\delta\rho^{\mathrm{visc}}_D\propto -H^2/T^5, \delta\rho^{\mathrm{pl}}_{D}\propto -H^2/T, \delta\rho^{\mathrm{H}}_D\propto H^2/T^5, \delta\rho^{\mathrm{th}}_D\propto H^2T^3$. 

For a regime of parameters where $\omega_c\tau_{\mathrm{ee}}<1$ one can ignore field dependence of shear viscosity coefficient and concentrate exclusively on the resonant collective mode contributions to drag. For example at the temperature of the drag peak $T\sim T_{\mathrm{pl}}$ the condition that  $\omega_c\tau_{\mathrm{ee}}<1$ is satisfied for all the fields up to the scale of plasmon frequency $\omega_c\lesssim \omega_{\mathrm{pl}}$. One could also easily estimate that $\alpha(T_\mathrm{pl})\sim\beta(T_{\mathrm{pl}})\sim 1/(k_Fd)$. In this regime drag resistance is given by Eq. \eqref{md-rho-mpl}, which for the FL parameters reads as 
\begin{equation}
\rho_D(H)\simeq \frac{\rho_Q}{(k_Fd)(R_ck_F)^2}\left(\frac{T}{E_F}\right)^3\propto T^3 H^2
\end{equation}
This result enables us to estimate the enhancement of the drag peak by magnetic field $\rho^{\mathrm{max}}_D(H)/\rho^{\mathrm{max}}_D\sim (k_Fd)(d/R_c)^2$. It is perhaps worth to mention that the expression for drag resistance in Eq. \eqref{md-rho-mpl} bears a close resemblance to recently obtained viscous magnetoresistance of a single electron layer subject to smooth inhomogeneous potential with long correlation radius \cite{AL-Viscous-MR}, in particular its inverse proportionality to viscosity as opposed to the linear proportionality of the zero-field resistance. This is not accidental as both have similar physical origin. Indeed, in the presence of the current, the external potential moves relative to the liquid and produces fluctuations of density in the electron liquid. The subsequent scattering of density fluctuations from the disorder potential produces a net resistive force. Thus intralayer resistivity can be understood in terms of the drag force between the electron liquid and the disorder potential. 

\subsection{Outlook}

In closing, we mention that within the same hydrodynamic approach we examined a question of whether Hall viscosity, which enters stress tensor as a transverse term, can render finite Hall drag resistance. However, upon a close inspection of various terms we do not find a nonvanishing contributions at least within main hydrodynamic approximation. It is still possible that higher gradient corrections to hydrodynamic equations of motion, due to formal expansion in Knudsen number, and an additional gradient terms, due to density dependence of kinetic coefficients, can lead to transverse correlations that generate Hall drag. In this work, we have not delved into an exhaustive analysis of these possibilities, which should be pursued in a separate study.

\subsection*{Acknowledgments}

We thank Anton Andreev for the interest and valuable comments at the early stages of this project, and we thank Igor Burmistrov for discussions on Hall viscosity. 
This work was supported by the National Science Foundation Grants No. DMR-1853048 (DAP) and DMR-1653661 (AL). This work was performed in part at Aspen 
Center for Physics, which is supported by National Science Foundation grant PHY-1607611

\appendix 

\section{Correlation function of fluctuating viscous stresses and thermal fluxes} \label{appendix-1}

Theory of hydrodynamic fluctuations was developed by Landau and Lifshitz \cite{LL-57} based on earlier ideas of Rytov \cite{Rytov} who also developed the same approach to electromagnetic fluctuations. 
The general idea of the method can be summarized as follows. Let us view the expressions for the stress tensor $\sigma_{ik}$ and heat flux $Q_i$ in the fluid
\begin{align}
&\sigma_{ik}=2\eta v_{ik}+(\zeta-\eta)\delta_{ik}\partial_lv_l+\eta_H(\varepsilon_{ijz}v_{kj}+\varepsilon_{kjz}v_{ij})+\varsigma_{ik} \label{stress-flux}\\ 
&Q_i=-\kappa \nabla_i T+g_i \label{thermal-flux}
\end{align}
as an equation of motion for a random variable $x_a$ subject to a random force $y_a$
\begin{equation}\label{x}
\dot{x}_a=-\sum_b \gamma_{ab}X_b+y_a. 
\end{equation}
where the first term in the right-hand-side describes relaxation with dissipative coefficient $\gamma$. The meaning of conjugated (or dual) variables $X_a$ is identified by entropy production 
\begin{equation}
\dot{S}=-\sum_a\dot{x}_aX_a.
\end{equation}
For the Gaussian distribution of $x_a$ the correlation function of $y_a$ is then given by 
\begin{equation}\label{y}
\langle y_a(t)y_b(t')\rangle=(\gamma_{ab}+\gamma_{ba})\delta(t-t')
\end{equation}
which signifies fluctuation-dissipation relation and Onsager symmetry principle.  

In this setup we want to identify $\dot{x}_a$ with either $\sigma_{ik}$ or $Q_i$ so that force $y_a$ should be identified with $\varsigma_{ik}$ or $g_i$ respectively. Even though $x_a$ represent a discrete set of variables we can think of partitioning system into small volumes $\Delta V$ and then taking a continuum limit $\Delta V\to0$ to get to fields $\sigma_{ik}$ and $Q_i$. To realize this plan we first calculate the rate of entropy production in the fluid 
\begin{equation}
\dot{S}=\int\left[\frac{\sigma_{ik}}{2T}\left(\frac{\partial v_i}{\partial x_k}+\frac{\partial v_k}{\partial x_i}\right)-\frac{Q_i}{T^2}\frac{\partial T}{\partial x_i}\right]\mathrm{d}V.
\end{equation}
Two points are important here. First is that $\dot{S}$ does not contain cross terms between viscous and thermal fluxes, which implies that their fluctuations are uncorrelated. The second point, is that the trace $\sigma_{ik}v_{ik}=\eta v_{ik}v_{ik}+(\zeta-\eta)(\dv \bm v)^2$ is independent of Hall viscosity $\eta_H$, which reinforces the fact of its non-dissipative nature. From this expression of $\dot{S}$ we conclude that if we identify $\dot{x}_a\to \sigma_{ik}$, then the conjugated variable is $X_a\to-(\Delta V/T)v_{ik}$. Conversely, if we identify $\dot{x}_a\to Q_i$ then $X_a\to (\Delta V/T^2)\partial_iT$. With this prescription we can write stress tensor \eqref{stress-flux} in the form of an equation of motion \eqref{x} as follows 
\begin{equation}
\sigma_{ik}=-\sum_{lm}\gamma_{ik,lm}\left(-\frac{\Delta V}{T}v_{lm}\right)+\varsigma_{ik}
\end{equation}
which enables us now to identify the corresponding dissipative tensor 
\begin{align}
\gamma_{ik,lm}=\frac{T}{\Delta V}[\eta(\delta_{ik}\delta_{lm}+\delta_{im}\delta_{kl})+(\zeta-\eta)\delta_{ik}\delta_{lm}\nonumber 
\\ +\eta_H(\varepsilon_{imz}\delta_{kl}+\varepsilon_{kmz}\delta_{il})]
\end{align}
that should enter the correlation function \eqref{y} for $\varsigma_{ik}$. Taking a continuum limit $\Delta V\to0$ implies $(T/\Delta V)\to T\delta(\bm{r}-\bm{r}')$. 
This way we arrive at Eq. \eqref{Lengevin-stress-corr-fun} from the main text. Hall viscosity piece of $\gamma_{ik,lm}$ drops out from the correlation function 
because of symmetrization with respect to pairs of indices while terms with $\eta_H$ are antisymmetric and cancel pairwise.    

In complete analogy,  we can present the heat flux \eqref{thermal-flux} in the form of Eq. \eqref{x} as follows 
\begin{equation}
Q_i=-\sum_{ik}\gamma_{ik}\left(\frac{\Delta V}{T^2}\partial_kT\right)+g_i
\end{equation} 
with
\begin{equation}
\gamma_{ik}=\frac{\kappa T^2}{\Delta V}\delta_{ik}
\end{equation}
After symmetrization and in the continuum limit this gives Eq. \eqref{Langevin-thermal-corr-fun} from the main text. In closing this section we mention that this approach can be generalized to the regime of quantum fluctuations. In that case few modifications are needed: first, is that thermal factor should be replaced by a proper distribution function of bosonic modes of given frequency $\omega$; second, is that kinetic coefficients become dispersive and only their real part enters the correlation function. For example, quantum thermal fluctuations are described by the correlator (a time Fourier transform):
\begin{equation}
\langle g_{i}(\bm{r},\omega)g_j(\bm{r}',-\omega)\rangle=T\omega\coth(\omega/2T)\Re\kappa(\omega)\delta(\bm{r}-\bm{r}'). 
\end{equation}  

\section{Asymptotic behavior of integrals}\label{appendix-2} 

The leading asymptotic behavior of $F_0(\beta)$ in Eq. \eqref{F-0} and $F_1(\alpha,\beta)$ in Eq. \eqref{F-1} can be captured as follows. For any finite and small $\beta$ the integrand of $F_0$ increases from $x=0$ to a certain maximum value at the point of $x_0\sim\ln(1/\beta)$, and then exponentially decays as $x\to\infty$. Taking into account this behavior and the fact that the peak is relatively sharp we can approximate $\beta^2 x^3\to\beta^2 x^3_0$ in the denominator of Eq. \eqref{F-0} since in the main logarithmic approximation we have to treat $\beta^2 x^3\sim\mathrm{e}^{-2x}\ll1$, so that 
\begin{align}\label{F0-approx}
&F_0(\beta)\approx\int\limits^{\infty}_{0}\frac{x^4\mathrm{d}x}{\beta^2 x^3_0\mathrm{e}^{2x}+1}=\nonumber \\ 
&-\frac{3}{4}\Li_{5}\left(-\frac{1}{\beta^2 x^3_0}\right)\approx \frac{1}{160}\ln^5\left[\frac{1}{\beta^2\ln^3(1/\beta)}\right].
\end{align} 
In the last approximate equality sign we retained only the leading log in the expansion of the polylogarithmic function $\Li_5(z)$ at its large argument, which also contains a series of logs of smaller powers. 

Function $F_1$ behaves qualitatively similar to $F_0$ and also admits representation in polylogarithmic functions. Indeed, the integrand of $F_1$ in Eq. \eqref{F-1} reaches maximum at a point $x_1$ which is parametrically close to $x_0$ with logarithmic accuracy $x_1\sim\ln(1/\beta)$. By approximating $\beta^2 x^3\to \beta^2x^3_1$ we find 
\begin{align}
&F_1(\alpha,\beta)\approx\int\limits^{\infty}_{0}\frac{3\beta^2x^3_1(1-2\alpha x^2)x^3\mathrm{d}x}{(\beta^2 x^3_1\mathrm{e}^{2x}+1)^2}=\nonumber \\ 
&-\frac{9}{8}\Li_3\left(-\frac{1}{\beta^2x^3_1}\right)-\frac{90\alpha}{8}\Li_5\left(-\frac{1}{\beta^2x^3_1}\right).
\end{align}

\end{document}